\documentstyle[12pt,epsfig]{article} 
\begin{document} 
\bibliographystyle{unsrt} 
 \begin{center} 
{\large\bf Modelization of Thermal Fluctuations in G Protein-Coupled Receptors} 
 
\vspace{0.2cm} 
{\underline{C. Pennetta$^1$}, V. Akimov$^1$, E.Alfinito$^1$,  
L. Reggiani$^1$,G. Gomila $^2$, G.Ferrari$^3$, L. Fumagalli$^3$ and M. Sampietro$^3$} 
 
\vspace{0.2cm} 
$^1$ {\it Dipartimento di Ingegneria dell'Innovazione,  
Universit\`a di Lecce and \\National Nanotechnology Laboratory-INFM,  
Via Arnesano, 73100 Lecce, Italy.} 
 
\vspace{0.2cm} 
$^2$ {\it Department d'Electronica and Research Centre for Bioelectronics 
and Nanobioscience, Universitat de Barcelona, C/ Josep Samitier 1-5, 08028 
Barcelona, Spain.} 
 
\vspace{0.2cm} 
$^3$ {\it Dipartimento di Elettronica ed Informazione, Politecnico di Milano,  
P.zza Leonardo da Vinci 32, 20133 Milano, Italy} 
 
\end{center}

\begin{abstract}  
We simulate the electrical properties of a device realized by a G protein   
coupled receptor (GPCR), embedded in its membrane and in contact with two   
metallic electrodes through which an external voltage is applied. To this   
purpose, recently, we have proposed a model based on a coarse graining   
description, which describes the protein as a network of elementary   
impedances. The network is built from the knowledge of the positions of  
the C$_{\alpha}$ atoms of the amino acids, which represent the nodes of the   
network. Since the elementary impedances are taken depending of the  
inter-nodes distance, the conformational change of the receptor induced by  
the capture of the ligand results in a variation of the network impedance.  
On the other hand, the fluctuations of the atomic positions due to thermal  
motion imply an impedance noise, whose level is crucial to the purpose of an  
electrical detection of the ligand capture by the GPCR. Here, in particular,  
we address this issue by presenting a computational study of the impedance  
noise due to thermal fluctuations of the atomic positions within a rhodopsin  
molecule. In our model, the C$_{\alpha}$ atoms are treated as independent,  
isotropic, harmonic oscillators, with amplitude depending on the temperature  
and on the position within the protein ($\alpha$-helix or loop). The relative   
fluctuation of the impedance is then calculated for different temperatures.   
\end{abstract}  

\section{Model and Results}  
\vspace*{-0.25cm}  
G protein-coupled receptors (GPCRs) constitute the largest family of   
trans-membrane receptors, with functions going from revealing light and   
smells to the individuation of drug and virus intruders \cite{superfamily}.   
For this reason, many efforts are devoted to the study of their properties   
\cite{superfamily}. In particular, we are interested to develop electronic   
nanobiosensors based on GPCRs. Actually, the detection of an electrical signal  
from hybrid nanodevices based on a single or few receptors and associated   
with the capture of the ligands, is a challenging goal, rich of potential   
applications \cite{joachim,elec_detec}. Here, our aim concerns with the   
calculation of the electrical properties of a device realized by a G protein   
coupled receptor \cite{superfamily}, embedded in its membrane and in contact   
with two metallic electrodes through which an external AC voltage is applied  
\cite{pen_fn04}. To this purpose, recently, we have proposed a model based on   
a coarse graining \cite{atilgan} description, which describes the protein as  
a network of elementary impedances \cite{pen_fn04}. The network is   
built  from the knowledge of the position of the C$_{\alpha}$ atoms of  
the amino acids \cite{pdb}, which are taken as the nodes of the network  
\cite{albert}. At least in the case of rhodopsin (photonic receptor) these   
positions are known by X-rays diffraction experiments \cite{pdb} for both,   
the basic and the most stable excited state (metarhodopsin) \cite{pdb}.   
Though these positions are generally unknown for the other receptors, a   
coarse graining, complex network approach offers the possibility of taking   
advantage of the common topology of the GPCR family \cite{superfamily}.   
In fact, all GPCRs share a seven-helices trans-membrane structure, where the   
seven $\alpha$-helices are interconnected by extracellular and intracellular   
loops \cite{superfamily,pdb}. Additionally, there are two terminal chains: an  
extracellular chain (N terminus) and an intracellular chain (C terminus)  
\cite{superfamily,pdb}. We assume that the amino acids interact electrically   
among them and that charge transfers between neighboring residues and/or   
changes of their electronic polarization \cite{song} affect these   
interactions \cite{pen_fn04}. Accordingly, a link is drawn between any pair   
of nodes neighboring in space within a given a distance, $d=2R_a$ ($R_a$   
electrical interaction radius) \cite{atilgan,albert} and an elementary   
impedance is associated with each link \cite{pen_fn04}. Moreover, two extra   
nodes can be introduced in the network, associated with the electrodes, which  
are linked to a given set of amino acids, depending on the particular   
geometry of the contacts in the real device. The elementary impedance is   
taken as the impedance of a RC parallel circuit (the most usual equivalent   
passive AC circuit) \cite{pen_fn04}. Precisely, by denoting with $Z_{i,j}$   
the impedance associated with the link between the $i$-th and $j$-th nodes,   
separated by a distance $l_{i,j}$, we take \cite{pen_fn04}:  
\vspace*{-0.25cm}  
\begin{equation}  
Z_{i,j}={l_{i,j}\over \pi (R_a^2 -l_{i,j}^2/4)}   
{1\over (\rho^{-1} + i \epsilon_{i,j}\epsilon_0\omega)}  
\label{eq:zijsmoth}  
\end{equation}    
\vspace*{-0.10cm}  
where $\omega$ is the frequency of the external voltage, $\rho$ the  
resistivity of the resistor, $\epsilon_0$ the vacuum permittivity  
and $\epsilon_{i,j}$ the relative dielectric constant of the capacitor \cite{pen_fn04}  
expressed in terms of the intrinsic polarizabilities $\alpha_i$ and $\alpha_j$  
of the corresponding amino acids \cite{song}. By taking the values  
$R_a=12.5$ \AA\ and $\rho=10^9$ $\Omega$m, we have found \cite{pen_fn04} that  
the conformational change of the receptor induced by the capture of the ligand  
(i.e. the transition rhodopsin $\rightarrow$ metarhodopsin) implies a  
significant variation of the impedance at all frequencies, and in particular  
we have found a variation of about 20 \% in the static value of $Re[Z]$   
\cite{pen_fn04}. On the other hand, the fluctuations of the atomic positions   
due to the thermal motion \cite{atilgan,parak,tirion} imply an impedance   
noise, whose level, in comparison with the impedance change due to variation   
of conformation and with the electrode/amplifier noise, is crucial to the   
purpose of an electrical detection of the ligand capture by the GPCR.   
Therefore, here we consider the effect of the thermal atomic motion on the   
electrical response to an external field of a rhodopsin molecule. To this   
purpose, we allow the nodes of the network (C$_\alpha$ atoms) to fluctuate   
around their equilibrium positions. For the sake of simplicity and to get a   
qualitative estimation, we describe the system of coupled oscillators as a   
set of independent, isotropic, harmonic oscillators. When the oscillators are  
in their ground state, their positions, $\vec r$, referred to the equilibrium  
ones, are distributed with a probability density:  
\vspace*{-0.2cm}  
\begin{equation}  
|\psi(\vec r)|^2 = {\bigl({M\omega_0 \over \pi \hbar}\bigr)}^{3/2}  
\exp \bigl[-{M\omega_0 \over \hbar}r^2 \bigr]={1\over (2\pi<x^2>)^{3/2}}  
\exp \bigl[-{1\over 2} {r^2 \over<x^2>} \bigr]  
\label{eq:prob}  
\end{equation}   
\vspace*{-0.1cm}  
where $M$ is the average mass of the amino acids, $\omega_0=\sqrt{\gamma/M}$   
the oscillator frequency, $\gamma$ the force constant and $<x^2>=1/3<r^2>$   
is the mean square displacement of the oscillator from its equilibrium   
position along the x-direction. If each oscillator is in contact with a   
thermal bath at temperature $T$, the value of $<x^2>$ at the equilibrium is:  
%
\begin{equation}  
<x^2> ={1\over 2} {k_B\theta \over \gamma}+ {k_B\theta \over 3 \gamma}  
{1\over \ \exp[\theta/T] -1}  
\label{eq:mean}  
\end{equation}   
%
where $\theta=\hbar\omega_0/k_B$. When $T\gg\theta$, Eq.~(\ref{eq:mean})   
simplifies in:  
\vspace*{0.3cm}
\begin{equation}  
<x^2>\approx k_{B}T/ 3 \gamma  
\label{eq:thigh}  
\end{equation}   
\vspace*{0.3cm}  
\begin{figure}  
\begin{center}
\includegraphics[height=.3\textheight]{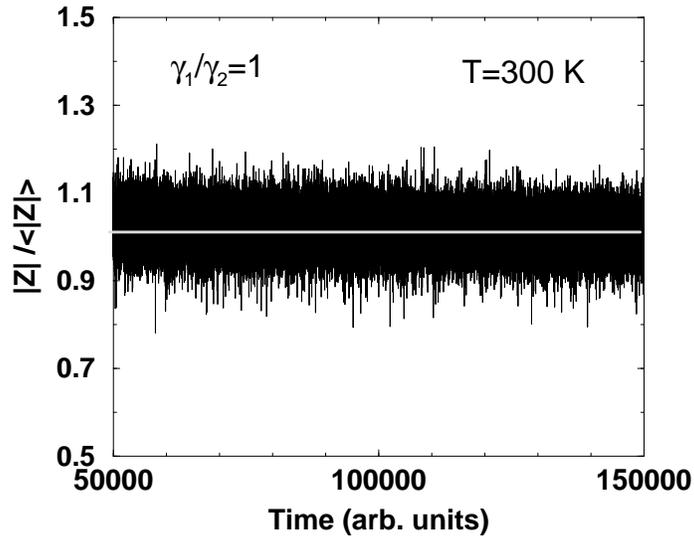}  
\caption{Simulation of the modulus of the network impedance, $|Z|$,   
versus time at $T=300$ K. The time is expressed in simulation steps while  
$|Z|$ has been normalized to its average value (shown by the gray line).} 
\end{center}
\end{figure}  
\vspace*{-1.4cm}  
\begin{figure}
\begin{center}  
\caption{Relative root mean square fluctuation of the impedance modulus as a  
function of the temperature (in K). The three set of data refer 
to simulations  
performed wit the ratio of the helix and loop force constants,  
$\gamma_1/\gamma_2$, equal, respectively, to $1$ (full circles), $50$  
(open squares) and $60$ (full diamonds). The dashed curves show the best-fit 
 with an exponential function.}  
\includegraphics[height=.3\textheight]{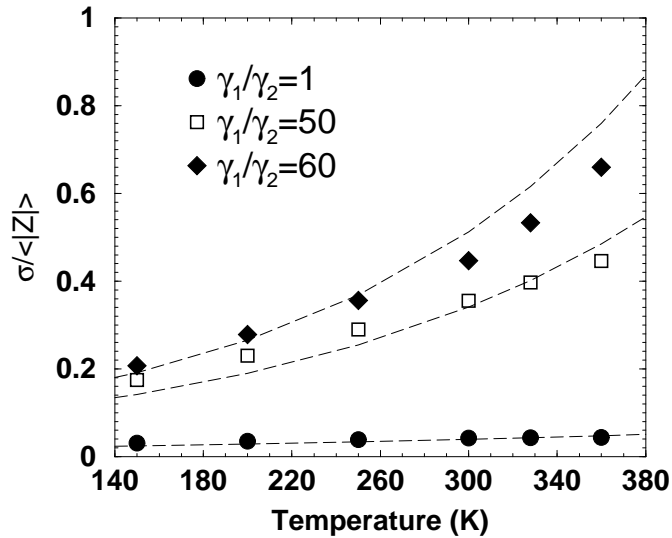}  
\end{center}
\end{figure} 

%
Of course, for an arbitrary temperature\ $T\ne 0$ the wave function of the   
oscillator is a superposition of several excited states and the density   
probability cannot be expressed in the simple form of Eq.~(\ref{eq:prob}).   
However, again for simplicity, we keep this expression and we account for 
the effect of the temperature by assuming in Eq.~(\ref{eq:prob}) $<x^2>$ 
given  by Eq.~(\ref{eq:thigh}). Moreover, as a first 
approximation, we take the mass   
$M$ and the force constant $\gamma$ equal for all the oscillators   
\cite{tirion}, and precisely: $M =\bar M=100$ Dalton,   
$\gamma=2.5$ KJ mole$^{-1}$ \AA $^{-2}$ \cite{atilgan}. This choice   
provides $\theta=12$ K. Thus, the condition $T\gg\theta$ is satisfied at   
room temperature. Figure 1 shows the results of simulations at $300$ K of   
the modulus of the network impedance, $|Z|$, versus time (the modulus has   
been normalized to its average value). On the other hand, it is well known   
\cite{superfamily,parak,frauenfelder} that loops and terminals are very   
flexible structures compared with the quite rigid $\alpha$-helices.   
Therefore, to overcome the crude approximation of a unique force constant   
for all the oscillators, we consider two different spring elastic constants,   
$\gamma_1$ and $\gamma_2$, for oscillators belonging to the $\alpha$-helices   
and to loops/terminals, respectively, with $\gamma_1 > \gamma_2$. Figure 2   
shows the relative root mean square fluctuation of the modulus of the  
impedance, i.e. the root mean square fluctuation, $\sigma $, normalized to  
the average value of $|Z|$, as a function of the bath temperature. The  
three sets of data report the results of simulations performed by taking the 
ratio, $F=\gamma_1/\gamma_2$, equal $1$ (full circles), $50$ (open squares) and $60$   
(full diamonds). In all cases the values of $\gamma_1$ and $\gamma_2$ have   
been chosen to keep constant the average value of the force constant,   
$\bar{\gamma}=(1/N) \sum \gamma_i$, where the sum is performed over the whole number 
$N$ of considered oscillators. The dashed curves in this figure represent the  
best-fit with exponential functions. We can see that once the higher flexibility  
of loops and terminals is accounted for, the relative fluctuation of the impedance  
increases significantly and it becomes strongly sensitive to the temperature.  
>From this study we can conclude that a careful modelization of thermal motion   
\cite{parak,frauenfelder} is necessary to provide reliable estimates of the   
relative fluctuation of the impedance.

{\large\bf Acknowledgments}

This work has been performed within the SPOT NOSED project   
IST-2001-38899 of EC. Partial support from the cofin-03 project   
``Modelli e misure di rumore in nanostrutture'' financed by Italian   
MIUR is also acknowledged. Authors thank E. Pajot-Augy, R. Salesse and J. Minic  
(INRA, Jouy en Josas, France) for helpful discussions.   
 
\bibliography{flu_gpcr} 

\begin{thebibliography}{10}

\bibitem{superfamily}
R.~J. Lefkowitz.
\newblock The superfamily of heptahelical receptors.
\newblock {\em Nature Cell Biology}, 2:E133--E136, 2000.

\bibitem{joachim}
C.~Joachim, J.K. Gimzewski, and A.Aviram.
\newblock Electronics using hybrid-molecular and mono-molecular devices.
\newblock {\em Nature}, 408:541--548, 2000.

\bibitem{elec_detec}
F.~Patolsky, G.~Zheng, O.~Hayden, M.~Lakadamyali, X.~Zhuang, and C.~M. Lieber.
\newblock Electrical detection of single viruses.
\newblock {\em PNAS}, 101:14017--14022, 2004.

\bibitem{pen_fn04}
C.~Pennetta, V.~Akimov, E.~Alfinito, L.~Reggiani, and G.~Gomila.
\newblock Fluctuations of complex networks: Electrical properties of single
  protein nanodevices.
\newblock In J.~M. Smulko, Y.~Blanter, M.~I. Dykman, and L.~B. Kish, editors,
  {\em Noise and Information in Nanoelectronics, Sensors and Standards II},
  number 5472 in Proceedings of SPIE, pages 172--182, Bellingham, 2004. Int.
  Soc. Opt. Eng.

\bibitem{atilgan}
A.~R. Atilgan, S.~R. Durell, R.~L. Jernigan, M.~C. Demirel, O.~Keskin, and
  I.~Bahar.
\newblock Anisotropy of fluctuation dynamics of proteins with an elestic
  network model.
\newblock {\em Biophys. J.}, 80:505--515, 2001.

\bibitem{pdb}
Research~Collaboratory for Structural~Bioinformatics.
\newblock {\em Protein data bank}.
\newblock State University of New Jersey, http://www.rcsb.org/pdb, 1.

\bibitem{albert}
R.~Albert and A.~L. Barabasi.
\newblock Statistical mechanics of complex networks.
\newblock {\em Rev. Mod. Phys.}, 74:47--97, 2002.

\bibitem{song}
Xueyu Song.
\newblock An inhomogeneous model of protein dielectric properties: intrinsic
  polarizabilities of amino acids.
\newblock {\em J. Chem. Phys.}, 116:9359--9383, 2002.

\bibitem{parak}
F.~G. Parak.
\newblock Physical aspects of protein dynamics.
\newblock {\em Rep. Prog. Phys.}, 66:103--129, 2003.

\bibitem{tirion}
M.~M. Tirion.
\newblock Large amplitude elastic motions in proteins from a single-parameter
  atomic analysis.
\newblock {\em Phys. Rev. Letl.}, 77:1905--1908, 1996.

\bibitem{frauenfelder}
P.~W. Fenimore, H.~Frauenfelder, and R.~D. Young.
\newblock Proteins as paradigms complex systems.
\newblock In S.~M. Bezrukov, H.~Frauenfelder, and F.~Moss, editors, {\em
  Fluctuations and Noise in Biological, Biophysical and Biomedical Systems},
  number 5110 in Proceedings of SPIE, pages 1--9, Bellingham, 2003. Int. Soc.
  Opt. Eng.

\end{thebibliography}
  
\end{document}